\documentclass[%
 reprint,
 amsmath,amssymb,
 aps,
]{revtex4-2}
\usepackage{float}
\usepackage{graphicx}
\usepackage{dcolumn}
\usepackage{bm}
\usepackage{multirow}
\usepackage{hyperref}

\hypersetup{
colorlinks=true,
linkcolor=blue,
citecolor=blue,
urlcolor=blue
}

\begin{document}


\title{Zinc dialkyldithiophosphates adsorption and dissociation on ferrous substrates:\\an \textit{ab initio} study}

\author{Francesca Benini}
\author{Paolo Restuccia}%
\author{Maria Clelia Righi}
 \email{clelia.righi@unibo.it}
\affiliation{%
 Physics and Astronomy Department, University of Bologna,\\
 Viale Berti Pichat 6/2, 40137, Bologna (Italy)
}%


\begin{abstract}
Zinc dialkyldithiophosphates (ZDDPs) have been commonly used as anti-wear additives in the automotive  industry for the past 80 years. The morphology, composition and structure of the ZDDPs phosphate-based  tribofilm, which is essential for its lubricant functioning, have been widely studied experimentally. However, despite their widespread use, a general agreement on their primary functioning mechanism is still lacking. The morphology and composition of the ZDDPs phosphate-based tribofilm have been widely studied experimentally, but the formation process and the relevant driving forces are still largely debated. In particular, it is unclear whether the stress-induced molecular dissociation occurs in the bulk oil or on the substrate. In this work, we employ \textit{ab initio} density-functional theory simulations to compare ZDDP fragmentation in vacuum and over a reactive substrate, considering the effects of surface oxidation on the dissociation path. Our results show that the molecular dissociation is highly endothermic in the absence of a supporting substrate, while in the presence of an iron substrate it becomes highly energetically favoured. Moreover, the presence of the substrate changes the reaction path. At the same time, surface oxidation reduces the molecule-substrate interaction. These findings provide valuable insights into the early stages of the formation of phosphate-based tribofilms.
\end{abstract}

\maketitle


\section{\label{sec:intro}Introduction}
After more than 80 years since their discovery, zinc dialkyldithiophosphates (ZDDPs) are still the most used anti-wear additives in the automotive industry~\cite{Spikes2004,barnes2001review,nicholls2005review}. Indeed, this class of materials is known for their capability to prevent wear, especially on steel~\cite{gosvami2015mechanisms,neville2007compatibility,martin1999antiwear}, thanks to the formation of a glassy phosphate-based tribofilm~\cite{Zhang2016, Spikes2004}.

Many details on the ZDDP tribofilm chemical composition and  morphology are known thanks to the great effort that has been put experimentally into its characterization\cite{zhang2020mechanochemistry, parsaeian2017new,Zhang2016,topolovec2007film,ueda2019crystallinity,Fang2023,dorgham2018transient,heuberger2007xps}: 
the film, typically around 80-150 nm thick and composed of flat-topped pads~\cite{topolovec2007film}, consists of a superficial layer of zinc and iron polyphosphate and/or phosphate; closer to the steel surface it presents a sulfur-rich layer~\cite{Spikes2004,zhang2020mechanochemistry,dorgham2018transient}; finally, although believed amorphous for years, it was recently found that ZDDP tribofilms hide a nanocrystalline structure, promoted by prolonged rubbing, which improves their durability.~\cite{ueda2019crystallinity}

Concerning the tribofilm formation process, there is a common agreement regarding its main steps~\cite{ueda2019crystallinity,heuberger2007xps}: first, surface adsorption and the initial formation of iron disulphide take place; secondly, the alkyl group transfers from oxygen to sulfur, converting dialkylthiophosphate to dithionyl phosphate; thirdly, intermolecular reactions enable phosphates to polymerize to build polyphosphate chains, the building blocks for the tribofilm; and finally, the amorphous film forms, which then becomes crystalline within a depolymerization process enhanced by prolonged rubbing at elevated temperatures~\cite{ueda2019crystallinity}. 

However, despite such general consensus, there are still matters of debate, for instance concerning what specifically drives ZDDP dissociation and the subsequent film formation, as well as the exact fragmentation path followed by the molecule and the role played by the substrate in such process~\cite{Sukhomlinov2021,Fang2023,dorgham2018transient}. Concerning the latter, some authors proposed that the substrate hardness might affect tribofilm quality~\cite{gosvami2015mechanisms, Kumar2020,mosey2005molecular}.
Indeed, the appearance of a protective film in tribological conditions is a common phenomenon, and many factors might affect its formation. This is why a deep understanding of the different roles played by various physical quantities, such as compressive and shear stress, temperature, and the interaction with the substrate, is crucial to design new lubricant additives~\cite{Chen2022} and unveil the functionality of commercial ones~\cite{Peeters2020}. These contributions are difficult to isolate experimentally in actual tribological conditions since they are all at play simultaneously at the nano-asperity contact.

Among all of these quantities, it has been suggested that shear stress within the base oil might be the driving force for ZDDP tribofilm formation. This hypothesis, described through the Stress-Promoted Thermal Activation model~\cite{Zhang2016,spikes2018stress,zhang2020mechanochemistry,Fang2023}, has been strengthened by experimental evidence of ZDDP tribofilm formation on a variety of substrates, ranging from metals~\cite{suominen2000electroless, mee1988study} to ceramics~\cite{sheasby1996comparison, mingwu1993tribological,haque2007non}, DLC coatings~\cite{haque2008study} and silicon~\cite{gosvami2015mechanisms}, suggesting that the molecule-substrate chemical interaction plays a minor role in the phenomenon. However, recent studies emphasized the role of reactive substrates, like the ones exposed during tribological experiments, as the primary driving force in the dissociation and tribofilm formation of commercial lubricant additives~\cite{Losi2021,Peeters2021,Chen2022}.

Finally, despite the widespread use of ZDDP additives, there is a scarcity of theoretical atomistic simulations that can provide insights into the ZDDP tribofilm formation process. In particular, the fundamental steps of the chemical reaction paths leading to the ZDDP dissociation, inaccessible from experiments, and the possible role of the molecule-substrate interaction are still missing. Indeed, none of the limited numbers of computational works concerning ZDDP takes into account the presence of a substrate~\cite{ayestaran2021mechanochemistry, Mosey2003,Sukhomlinov2021,Sukhomlinov2022,mosey2004quantum}.  
In this work, we employ density-functional theory (DFT) to investigate ZDDP fragmentation on iron and oxygen-passivated iron substrates relevant to many technological applications. We study the dissociation path followed by the ZDDP molecule when bonds are stretched in vacuum to model the situation in which stresses arise in chemically non-reactive base oils and compare it with molecule fragmentation on ferrous substrates. In this way, we identified the most favorable dissociation path and analysed the effect of substrate passivation on ZDDP adsorption.

\section{\label{sec:comp}Methods}

\begin{figure}[htpb]
\includegraphics[width=\columnwidth]{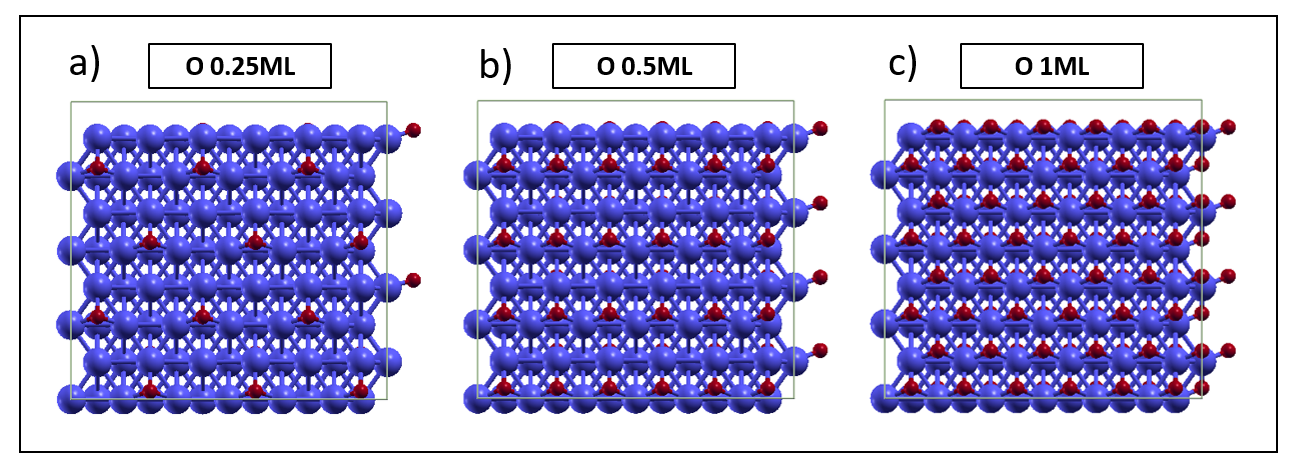}
\caption{\label{fig:sub}The oxidised substrates modelled in this study: 0.25ML (panel a), 0.5ML (panel b) and 1ML O-passivated Fe (panel c). From now on, Fe will be represented in blue and O in red.}
\end{figure}

All calculations were performed using spin-polarized Density Functional Theory (DFT) as implemented in version 6.8 of the Quantum Espresso suite~\cite{Giannozzi2009, Giannozzi2017, Giannozzi2020}. The exchange-correlation functional was described using the generalized gradient approximation (GGA) within the Perdew-Burke-Ernzerhof (PBE) parametrization~\cite{Perdew1996}. 
The kinetic energy cutoff for the wave functions was set to 40 Ry, with a cutoff for charge densities of 320 Ry. A gaussian smearing with a 0.001 Ry width was included to better describe occupations around the Fermi level. The electronic configuration of atoms was described using ultrasoft pseudopotentials in the RRKJ parametrization~\cite{rkkj}. The default energy and forces convergence threshold criteria (i.e., 10$^{-4}$ Ry and 10$^{-3}$ Ry/Bohr, respectively) were used to optimize the structural geometries. The geometrical configuration and computational setup we used to describe the ZDDP molecule, was already tested by our group in a previous work~\cite{Peeters2022}. In particular, the two lateral alkyl chains have been reduced to methyl groups to minimize the computational cost. 
Here we study the interaction of the ZDDP molecule with four different substrates: a clean Fe(110) four-layers 
thick slab, and, by passivating it with oxygen atoms on both sides, oxidised Fe slabs with 0.25 ML (Figure~\ref{fig:sub}a), 0.5 ML (Figure~\ref{fig:sub}b) and, 1 ML (Figure~\ref{fig:sub}c) O-coverages , respectively.  

To obtain an accurate description of the most effective dissociative path and of the role played by the substrate, the fragmentation of the molecule was studied by bond stretching~\cite{ta2023atomistic} comparing the results obtained in vacuum and in the presence of the clean Fe substrate. 
To carry out this task in vacuum, we placed the molecule vertically in a large supercell, and performed a sequence of relax calculations keeping the central Zn atom fixed while displacing firstly a P, then O, and finally a C atom of 0.2~\r{A} respectively along the $z$ direction at each step, reaching a total displacement of 5~\r{A}. The $z$ coordinates of the displaced atoms were kept fixed, and the optimized coordinates of each step were used as the initial configuration for the next step. An analogous procedure was performed for the case of the molecule on the iron slab. We considered the optimized molecule adsorbed over the substrate as a starting configuration. Then, we stretched the molecule by displacing first P, then O and finally a C atom along the direction parallel to the surface, namely the $x$ direction, keeping their $x$ coordinate fixed together with the in-plane coordinates of the Zn atom. In this case, we reached a total displacement of 3~\r{A}, which was enough to see the complete fragmentation of the molecule. We could then compute the reaction energies and force components acting on the fixed atoms at each step along the $z$ ($x$) direction for the stretching in the vacuum (Fe substrate) case. We expressed the obtained forces as pressures dividing them by the in-plane area of the cell employed for the molecule in vacuum (2.76 nm$^2$) to allow for a rough estimation of the shear stress that should be applied to obtain the corresponding molecular deformation.

To study the adsorption of the ZDDP molecule on surfaces, we employed large 6 $\times$ 4 orthorhombic supercells containing 48 atoms per layer to avoid undesired lateral interaction among the molecule replicas. Moreover, a reasonable amount of void (at least 10 \r{A}) was added to each cell to guarantee a vertical distance among replicas. The resulting supercell dimensions were 17.10 \r{A} $\times$ 16.12 \r{A} $\times$ 30.78 \r{A}. The latter was obtained after optimizing a smaller 2 $\times$ 2 cell. Considering the large dimensions of the supercell, we used a $\Gamma$ point sampling of the Brillouin zone. We computed adsorption energies as follows:

\begin{equation}
    E_{ads} = E_{tot} - (E_{surf} + E_{adsorbate})
    \label{eads}
\end{equation}

where $E_{tot}$ is the energy of the optimized system composed of the substrate and the adsorbed molecule/atom, while $E_{surf}$ ($E_{adsorbate}$) is the energy of the isolated surface (adsorbate).

To study the ZDDP dissociation over iron substrates, we used the molecular fragments identified by the bond stretching study. For each of them, we identified the most favourable binding site and molecular orientation  employing Xsorb~\cite{Xsorb} and computing the dissociation energy $E_{frag}$ as follows:

\begin{equation}
    E_{frag} = \sum_{n=1}^N E_{F_n} - (n \times E_{surf} + E_{mol})
    \label{efrag}
\end{equation}

where $E_{F_n}$ represents the total energy of the most stable systems composed of the surface and each fragment alone at a time, and $E_{surf}$ ($E_{mol}$) represents the energy of the isolated surface (molecule). The value of $n$ in the formula depends on how many fragments the molecule is cut into. In our case, $n$ = 2, 3 and 4, respectively.

We further studied the ZDDP dissociative path looking at what happens at the interface of two Fe (or 0.25 ML and 1 ML oxidised Fe) surfaces in the presence of 1 GPa external load. To achieve this task, we realized uniaxial compressions in which we reduced the vertical size of the cell and allowed the ZDDP molecule to interact with one side of the slab and its own replica to simulate the interface. This approach induced molecular dissociation and gave us insights into the bond breaking that takes place at the interface under load.

\section{\label{sec:res}Results and discussion}
Here we report the results of our simulations: first, we analyse the effect of iron oxidation on the adsorption of molecular ZDDP. Secondly, we compare ZDDP dissociation in vacuum and in the presence of the clean substrate. We consider that the dissociation in vacuum is comparable to the one within the base oil, due to the shallow interaction between the lubricant additive and the hydrocarbon molecule of the oil. We then look at the effect of substrate oxidation on the dissociation energies, and finally consider how the dissociation proceeds when the molecule is set at the interface between two facing substrates.  

Modeling bulk Fe oxides, like hematite, is extremely computationally demanding, and it requires an exquisite fine-tuning of the computational parameters, thus we modelled the effect of iron oxidation using iron slabs passivated by oxygen atoms, as done in previous studies~\cite{Peeters2020,Losi2021,Peeters2022}. 

\subsection{\label{sec:ads}ZDDP adsorption}
The first important step in the formation of ZDDP tribofilms is the adsorption of the molecules on the substrate~\cite{Spikes2004,zhang2020mechanochemistry, heuberger2007xps}. In addition to the clean iron substrate we simulated three different degrees of surface O-passivation, namely 0.25 ML, 0.5 ML and 1 ML, shown in Figure~\ref{fig:sub}.

\begin{figure}[htpb]
\includegraphics[width=\columnwidth]{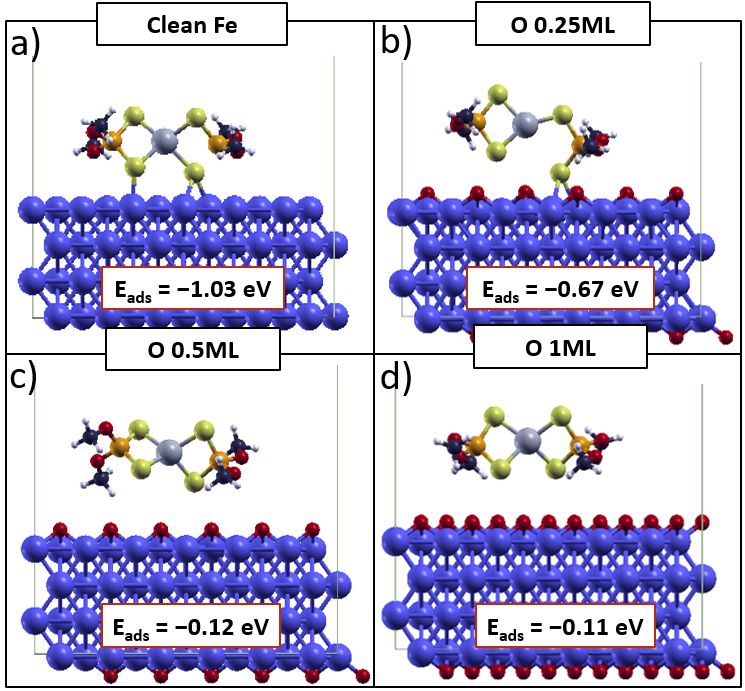}
\caption{\label{fig:full_mol_sub}ZDDP adsorption on a) clean Fe and O passivated Fe with three different O coverages: b) 0.25ML, c) 0.5ML and d) 1ML respectively. Adsorption energies are also reported. From now on, P, Zn, C, H and S atoms are represented in orange, grey, black, white and yellow, respectively.}
\end{figure}

After fully optimizing the geometry of the adsorbed molecule/substrate system, $E_{ads}$ was computed following Eq.~\ref{eads}. The optimized configurations, alongside the corresponding adsorption energies, are reported in Fig.~\ref{fig:full_mol_sub}. For all the considered systems, the adsorption energies are negative, indicating that the molecular adsorption is energetically favourable. This is in agreement with the experimental evidence that ZDDP undergoes chemisorption and thermal decomposition~\cite{Spikes2004, dacre1982adsorption}. However, oxygen passivation significantly reduces the interaction of the ZDDP molecule with Fe. In particular, $E_{ads}$ decreases by about one order of magnitude, from $-$1.03 eV for the bare iron substrate to $-$0.12 and $-$0.11 eV for the half and fully passivated substrates. The 0.25 ML passivated substrate maintains a mild reactivity, with $E_{ads}$= $-$0.67 eV, as the ZDDP undergoes structural deformations leading to bonds breaking and S interaction with iron atoms (Fig.~\ref{fig:full_mol_sub}b).

\subsection{\label{sec:stretch}Shear-induced dissociation of ZDDP in the absence and in the presence of a substrate}

\begin{figure*}[htpb]
\includegraphics[width=\textwidth]{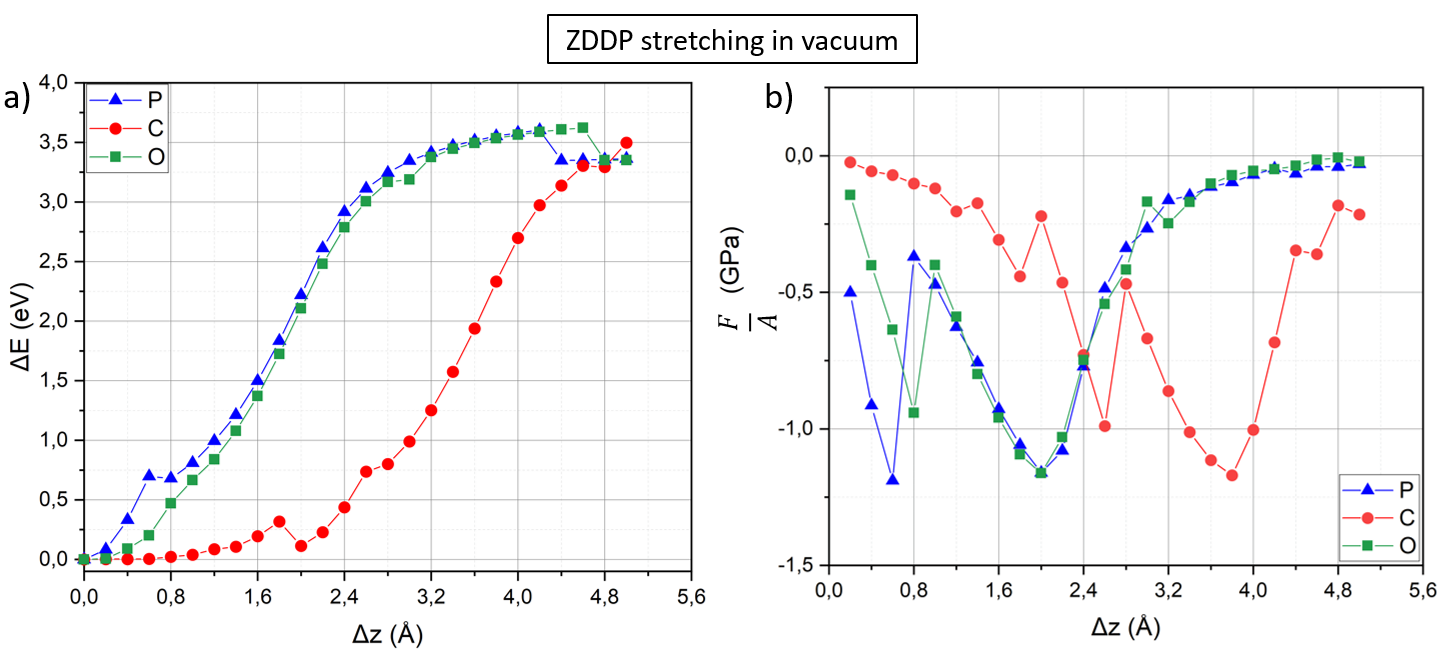}
\caption{\label{fig:stretch}Energy variation as a function of atom displacement along the $z$ direction (panel a) alongside force per unit area on P, C and O for the ZDDP molecule stretching in vacuum (panel b). $F$ is the value of the force acting on P, C and O whereas $A$ is the in-plane area of the cell. A positive (negative) value of $F$ indicates a force with the same (opposite) direction of the $z$ axis in the reference system.}
\end{figure*}

The results of the bond-stretching procedure applied in vacuum are shown in Fig.~\ref{fig:stretch}. In panel a) the energy variation as a function of the atom displacement along the $z$ direction are displayed for all three pulling atoms, whereas in panel b) the forces per unit area acting on P, C and O are shown. In all the three considered cases, the energy increases during the atom pulling until a plateau is reached at a fragments distance where the interaction is negligible. Independently from the pulling atom, the detachment of the S$_2$P(OCH$_3$)$_2$ fragment has an associated energy cost of about 3.5 eV. Alongside the energy, the force divided by the in-plane cell area acting on the pulled atoms, provides a consistent picture with a maximum increase of the resistant forces per unit area when bond breaking occurs, of about 1.2 GPa.

\begin{figure}[htpb]
\includegraphics[width=\columnwidth]{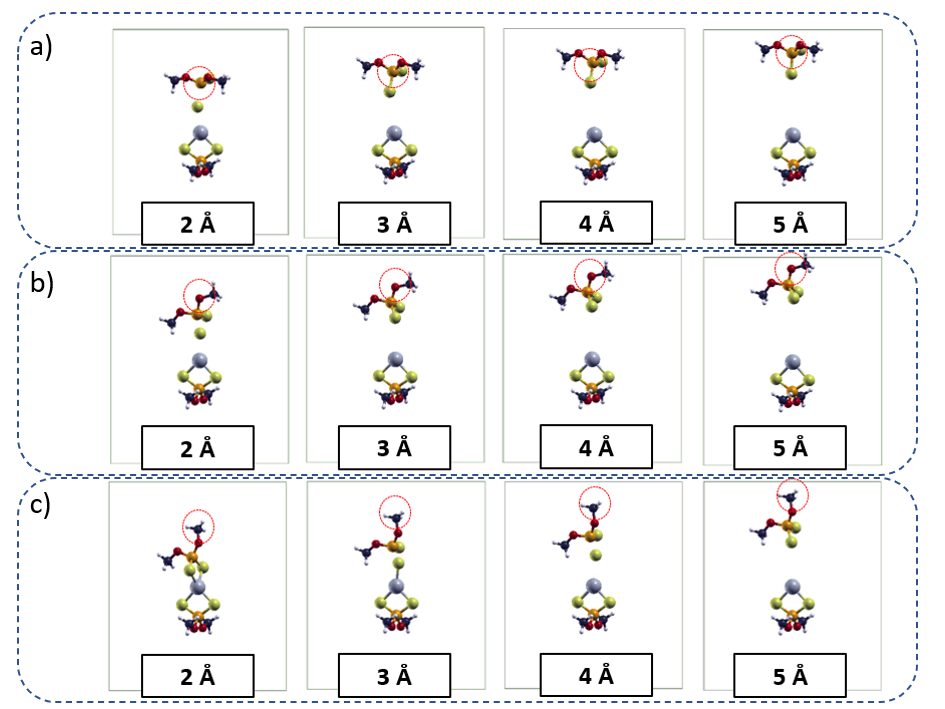}
\caption{\label{fig:stretch2}Configurations resulting from pulling, respectively, a P atom (panel a), an O atom (panel b) and a C atom (panel c). The atom pulled is highlighted with a dotted red circle and the displacement along $z$ is reported.}
\end{figure}

The geometry configurations along the stretching bond procedure are shown in Fig.~\ref{fig:stretch2}. Confirming what previously obtained in Ref.~\citenum{Peeters2022}, we find that, whatever atom is pulled, the most favourable fragmentation pattern involves breaking the Zn-S bonds. Therefore, we found that in vacuum or inside the lubricating oil, when shear stress is applied to the molecule, the Zn-S bonds are the less energetically demanding to break, as also reported in the literature.~\cite{tuszynski2002tribochemical,Zhang2016}

\begin{figure*}[htpb]
\includegraphics[width=\textwidth]{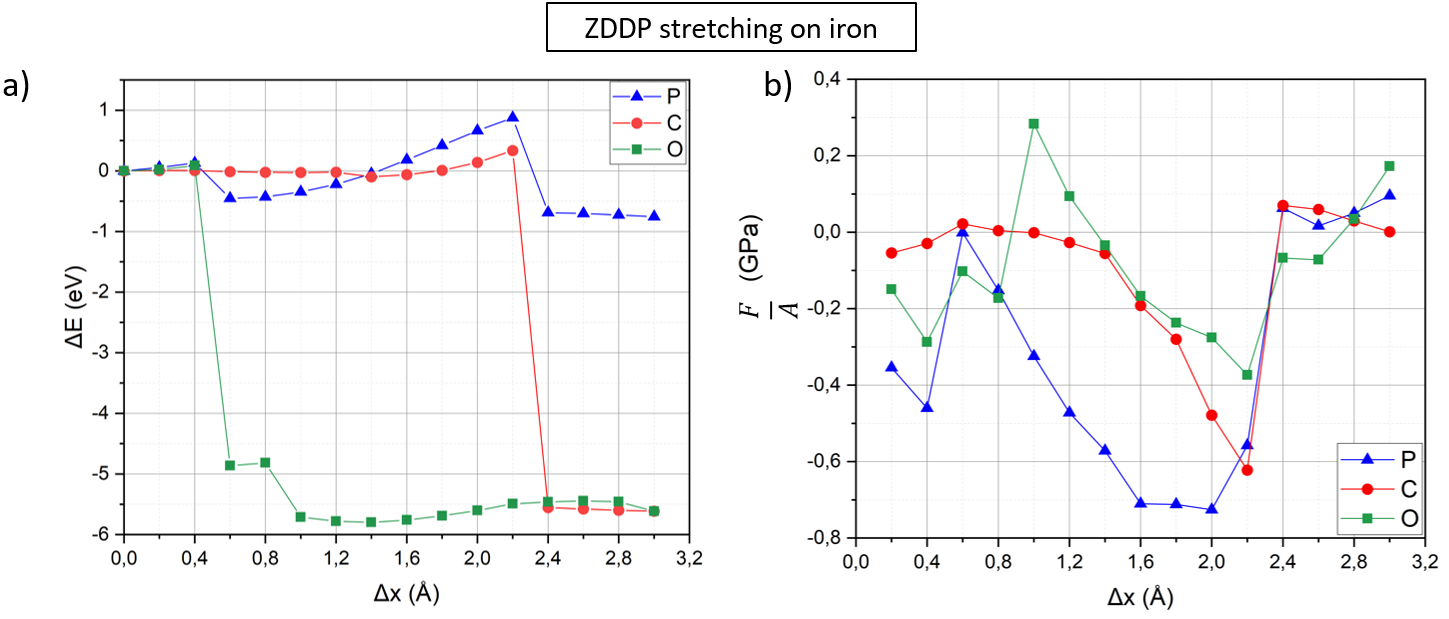}
\caption{\label{fig:forces_vacuum}Energy variation as a function of atom displacement along the $x$ direction (panel a) alongside force per unit area on P, C and O for the ZDDP molecule stretching on an iron substrate (panel b). $F$ is the absolute value of the force acting on P, C and O whereas $A$ is the in-plane area of the cell employed for the molecule in vacuuum, so that values are comparable. A positive (negative) value of $F$ indicates a force with the same (opposite) direction of the $x$ axis in the reference system.}
\end{figure*}

\begin{figure}[htpb]
\includegraphics[width=\columnwidth]{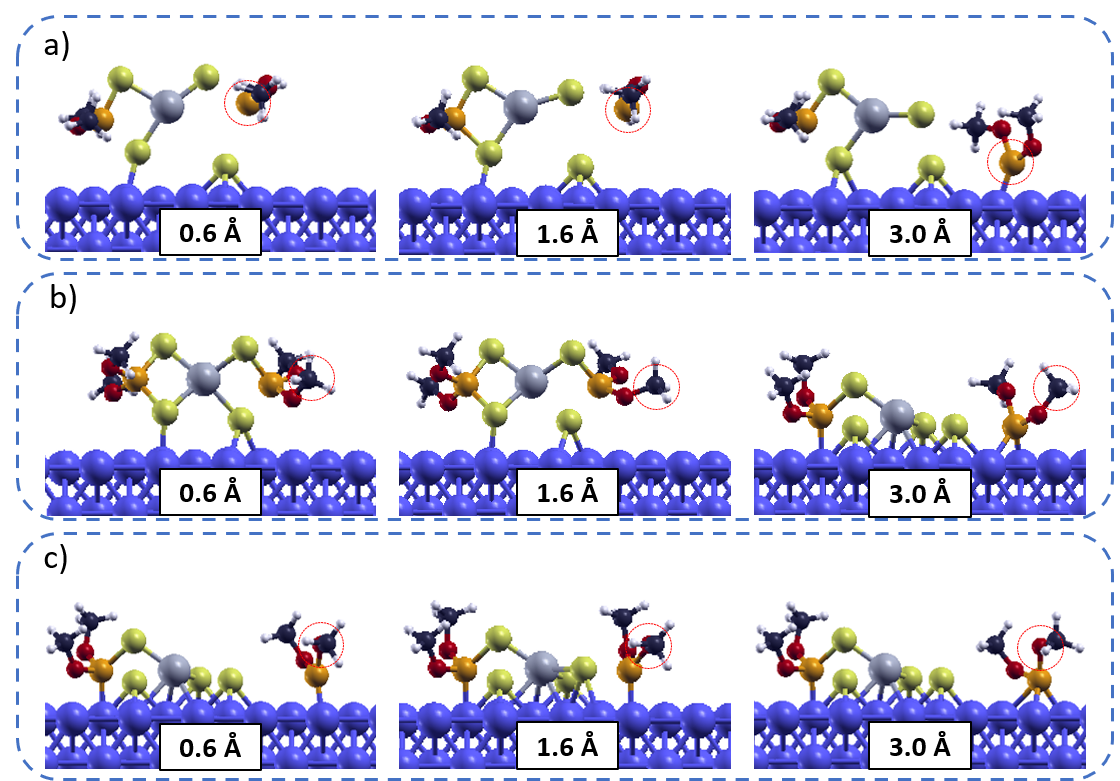}
\caption{\label{fig:stretchsub}ZDDP pulling on clean Fe. The atoms pulled, respectively P (panel a), C (panel b) and O (panel c) are highlighted with a dotted red circle. Atoms displacement is reported.}
\end{figure}

The results of the bond-stretching procedure on the iron substrate are summarized in Fig.~\ref{fig:forces_vacuum}.
Interestingly, the energy and the resistant stress in the presence of a substrate are much lower than those in the absence, indicating that the presence of the substrate makes molecular dissociation much easier than in the oil. Indeed, the dissociated fragments are stabilized by the substrate, as visible from the reaction energies that diminished significantly to around $-6$ eV for both C and O pulling (Fig.~\ref{fig:forces_vacuum}a). In the case of P, the reaction energy is significantly larger due to the shallower interaction of the Zn-S central unit compared to the other two cases (as shown in Fig.~\ref{fig:stretchsub}a).

Not only the energies and forces are dramatically changed by the presence of the substrate, but also the dissociation path is different. As shown in Fig.~\ref{fig:stretchsub}, the most favourable dissociation path involves the breaking of the P-S bonds. Therefore, the P(OCH$_3$)$_2$ groups are released and chemisorbed to the iron substrate, leaving the central Zn-S units and isolated S atoms adsorbed on iron. These results are in agreement with the experimental observation of an S-rich iron layer and phosphate layers within the tribofilm. A change in the dissociative path induced by the presence of a reactive substrate was found for MoDTC~\cite{Peeters2020}, another largely employed commercial lubricant additive.
Our results suggest that the substrate-mediated dissociation is much more probable than the suggested shear-induced fragmentation within the oil~\cite{Zhang2016} because the energies and forces involved are significantly lower. Moreover, the reactive substrate allowed to isolate the organophosphorus units from the central Zn-S ones, thus explaining the evidence of separate phosphate and zinc sulfide areas observed experimentally~\cite{Spikes2004, dorgham2018transient}.

\begin{figure}[htpb]
\includegraphics[width=\columnwidth]{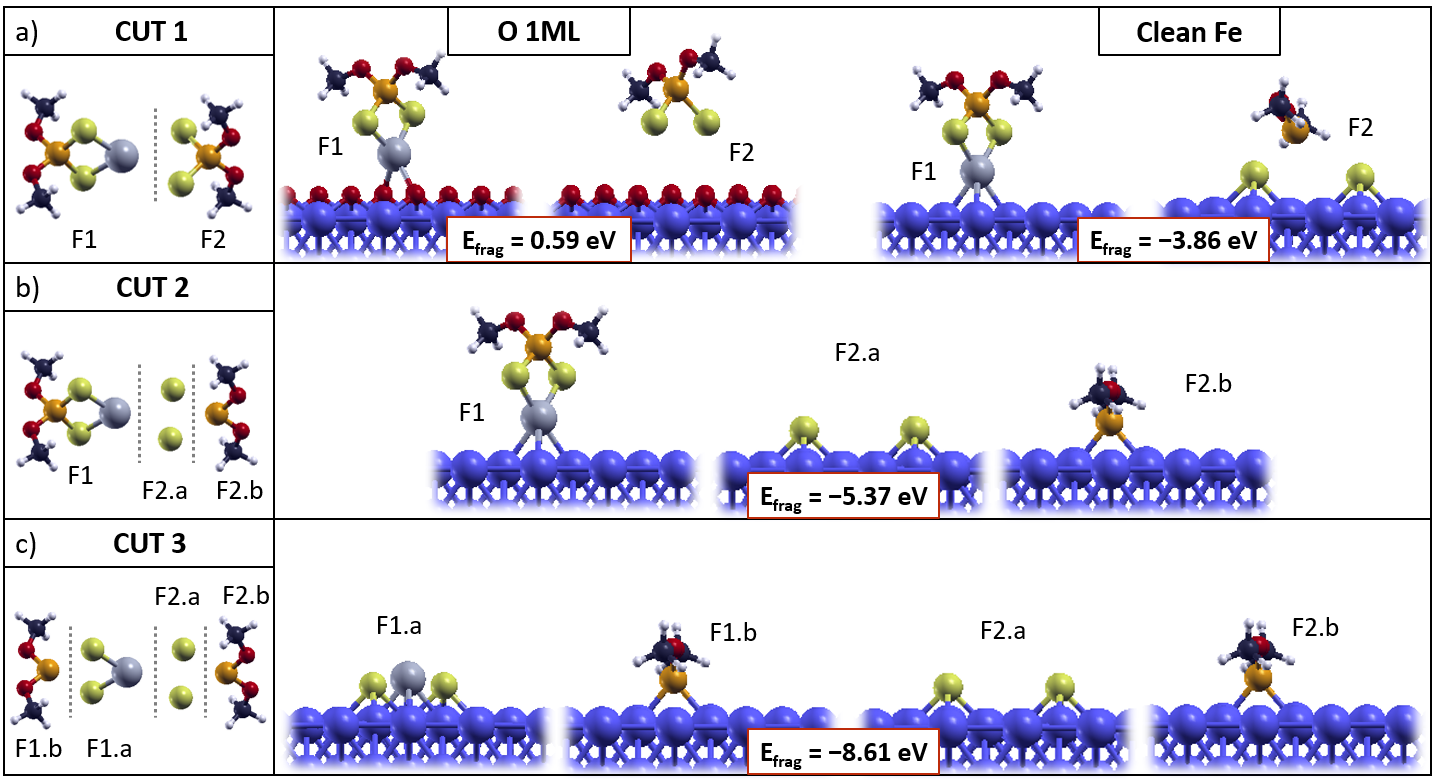}
 \caption{\label{fig:atoms_ads}ZDDP fragments adsorption following a two- (panel a), three- (panel b) and four-fragments (panel c) dissociation path. In panel a) both adsorption on 1 ML O-passivated Fe and clean Fe are reported, while panel b and c involve bare iron only.}
\end{figure}

Starting from the hypothesis that the molecule may dissociate in bulk oil, we first simulated the adsorption of the two fragments resulting from bond stretching in vacuum over a substrate (Fig.~\ref{fig:atoms_ads}). We performed a systematic study of their adsorption by employing Xsorb~\cite{Xsorb}, a computational tool developed in our group, to perform a configurational study testing different fragment orientations for each system. For the substrates, we take into account the two opposite conditions of O-passivation, namely a bare Fe(110) surface and a complete oxygen-passivated (i.e., 1 ML) Fe(110) substrate. We found the optimal adsorption geometries reported in Fig.~\ref{fig:atoms_ads}, together with the corresponding $E_{frag}$.

From this analysis shown in Fig.~\ref{fig:atoms_ads}a, we found that the adsorption of the Zn-terminated (F1) and the organophosphorus fragments (F2) over the O-passivated substrate is endothermic, with an associated energy cost of $+$0.59 eV. However, the same dissociation path on the bare iron substrate is energetically favorable, with $E_{frag}$ equal to $-$3.86 eV. These results confirm the effectiveness of O-passivation in hindering the ZDDP fragmentation completely. We highlight that the F2 adsorbed over the clean surface caused the S atoms (F2.a fragment) to further dissociate from the organophosphorus group (F2.b fragment). We tested this hypothesis by considering a three-fragments dissociation path (Fig.~\ref{fig:atoms_ads}b), which resulted in a further $E_{frag}$ decrease to $-$5.37 eV. This result suggests that organophosphorus groups prefer to stay isolated from the central units when adsorbed on iron, which is consistent with previous results of similar compounds on iron~\citenum{Righi-2015}. Complete fragmentation of the organophosphorus groups from the ZDDP (Fig.~\ref{fig:atoms_ads}c) resulted in the lowest $E_{frag}$ equal to $-$8.61 eV. We gained useful knowledge by studying the energeticis involved in dissociating ZDDP over a reactive substrate. Specifically, we found that it is energetically favorable to separate the P-terminated and Zn-S groups. This dissociation is the first step in creating the zinc sulfide and phosphate layers seen in previous experiments~\cite{Spikes2004, dorgham2018transient}.

\subsection{\label{sec:int}Load-induced ZDDP dissociation}
To verify the effects of load on molecular dissociation, we performed uniaxial compressions of the molecule confined at an iron interface in the presence of 1 GPa load, as explained in the Methods section. 

\begin{figure}[htpb]
\includegraphics[width=\columnwidth]{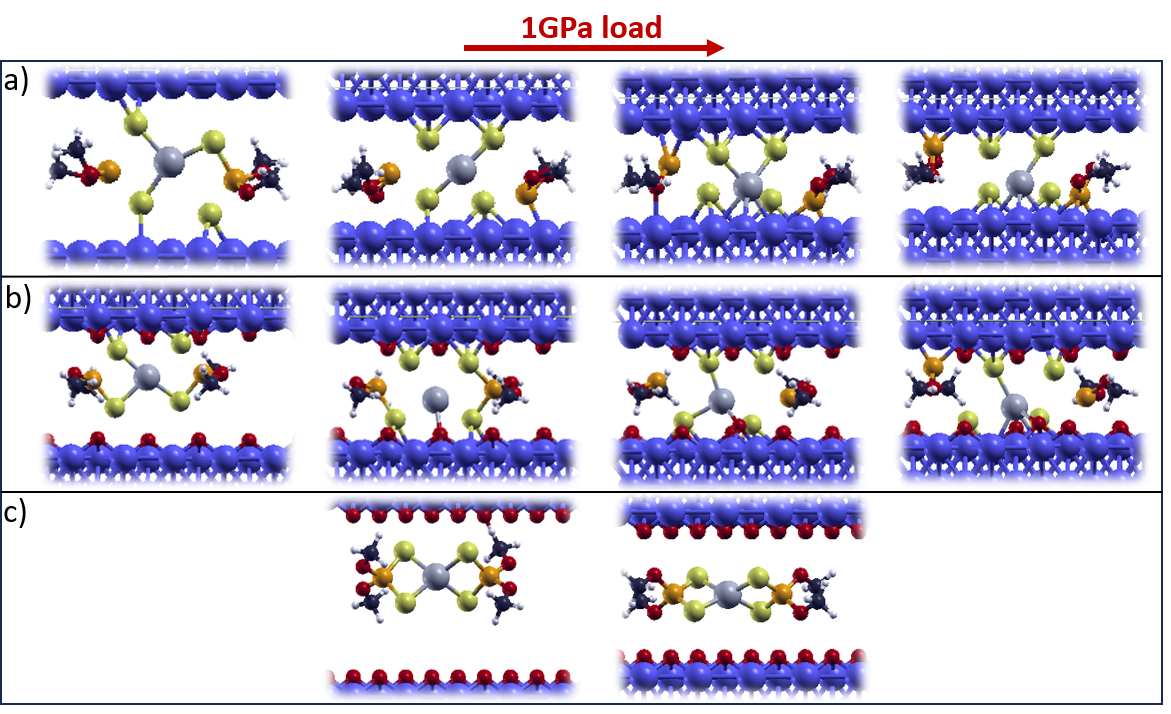}
\caption{\label{fig:vc} Uniaxial compression optimizations for ZDDP on clean Fe (panel a), 0.25 ML O passivated Fe (panel b) and 1 ML O-passivated Fe (panel c). From left to right, different snapshots while reaching the target 1 GPa pressure are reported.}
\end{figure}

In Fig.~\ref{fig:vc}, snapshots of the relaxation runs while reaching the target pressure are shown for the different oxidation levels. 
In Fig.~\ref{fig:vc}a and b, the snapshots relative to the clean and 0.25 ML O-passivated iron substrates, are shown. In both cases, the same dissociative paths determined from equilibrium calculations at the open surface is displayed. In particular, both organophosphorus groups separate from the ZDDP molecule, leaving the Zn-S central units completely isolated. These insights gained by dissociative reactions are fascinating since no direct experimental evidence exists on the most favourable ZDDP fragmentation paths. For instance, organophosphorus units always need to be saturated due to the 5 valence electrons present in the P atom electronic configuration. Therefore, in the gas phase, these fragments attracted the S atoms, breaking the Zn-S bonds. On the contrary, the reactive substrate helps stabilize the P-based groups that leave the molecule and adsorb on the surface.

Concerning the O-passivated iron surface at 1 ML, there are no dissociative reactions even when the external pressure is applied, and the ZDDP molecule only appears to be stretched. Therefore, the passivation of the iron substrate substantially reduces the rate of molecular dissociation and at high O-coverages it impedes it.

\section{\label{sec:concl}Conclusions}

We employed DFT simulation to evaluate key processes involved in the early stages of ZDDP tribofilm formation, namely: molecular adsorption and its stress-induced fragmentation both in vacuum and on ferrous substrate with different level of oxidation.

We found that oxygen passivation significantly reduces the interaction of the ZDDP molecule with iron. In particular, the energy gain for adsorption decreases by about one order of magnitude when the Fe (110) surface is fully oxidised. Moreover, the ZDDP dissociation is energetically favoured over the clean iron surface and this process is impeded by full oxidation of the iron substrate.

We studied both uniaxial compression and stretching of ZDDP and we observed similar fragmentation paths for both cases, forming P-terminated and Zn-S groups, which serve as the building blocks for the zinc sulfide and phosphate layers detected in post-mortem analysis~\cite{Spikes2004, dorgham2018transient}. Our findings shed light on the initial stages of tribofilm formation highlighting the crucial role of the substrate in favouring molecular dissociation. ZDDP dissociation in oil is expected to be highly endothermic. However, it has sometimes been proposed as the key mechanism.

Overall, our results provide a first atomistic insight into ZDDP dissociation, which represents the first stage of film formation. In particular, while shear stress application was commonly believed to induce the formation of ZDDP tribofilms, we showed how the chemical interaction between the molecule and the surface and how it is influenced, for instance by oxygen, might indeed play a crucial role as well.

\begin{acknowledgments}
These results are part of the SLIDE project, which received funding from the European Research Council (ERC) under the European Union’s Horizon 2020 research and innovation program. (Grant Agreement No. 865633). The authors want to thank Dr. Margherita Marsili for fruitful discussions. 
\end{acknowledgments}


\bibliography{bibliography}

\end{document}